\documentclass[12pt]{iopart}

\usepackage{iopams}

\usepackage{natbib}
\usepackage{epsfig}
\usepackage{amssymb}

\def \d{{\rm d}}

\def \PP{{\;.}}

\begin{document}

\title{Scale-free statistics of plasticity-induced surface steps on KCl single crystals}

\author{Jan Schwerdtfeger$^1$, Edward Nadgorny$^{1,2}$, Frederic Madani-Grasset$^1$,
Vasileios Koutsos$^1$, Jane R. Blackford$^1$ and Michael
Zaiser$^{1,2}$\footnote[3]{Tel. +44-131-6505671; Fax
+44-131-6513470; e-mail M.Zaiser@ed.ac.uk}}

\address{$^1$ The University of Edinburgh, Institute for Materials and Processes,
The King's Buildings, Sanderson Building, Edinburgh EH9 3JL, UK\\
$^2$ Michigan Technological University, Physics Department,
Houghton, MI, 49931, USA}

\begin{abstract}
Experimental investigations of plastic flow have demonstrated
temporal intermittency as deformation proceeds in a series of
intermittent bursts with scale-free size distribution. In the
present investigation, a corresponding spatial intermittency is
demonstrated for plastic flow of KCl single crystals. Deformation
bursts lead to large surface steps with a height distribution that
is consistent with the distribution of strain increments in
deformation of micro-columns, and the energy distribution of
acoustic emission bursts observed in deformation of macroscopic
single crystal samples of a wide class of materials.
\end{abstract}

\small{Keywords: plasticity, fluctuations, defects, avalanches.}

\maketitle

% main text
\section{Introduction}

In recent years, numerous experimental investigations have
demonstrated temporal intermittency of plastic flow in crystalline
solids. Weiss and co-workers (Weiss and Grasso 1997, Miguel et al.
2001, Richeton et al. 2006) investigated the acoustic emission of
plastically deforming ice and metal single crystals and
demonstrated that the acoustic emission signal consists of a
series of intermittent bursts which are characterized by
scale-free distributions of the energy $E$ (amplitude square
integrated over the duration of a burst) and peak amplitude $A$ of
the acoustic signal, with probability density functions $p(E)
\propto E^{-\kappa_E}$ and $p(A) \propto A^{- \kappa_A}$ that are
well described as power laws with exponents $\kappa_E \approx 1.5$
and $\kappa_A \approx 2$, extending over up to 8 decades with no
apparent cut-off.

Recently, Dimiduk and co-workers (2006) confirmed the temporally
intermittent nature of plastic flow in metallic single crystals by
direct observation. They carried out compressive deformation of
microcolumns machined out of Ni single crystals. The elongation
vs. time curves observed under stress-controlled loading are
characterized by an intermittent sequence of deformation jumps,
with elongation increments $\Delta l$ that are characterized by a
scale-free distribution $p(\Delta l) \propto \Delta l^{-\kappa_l}$
where, again, $\kappa_l \approx 1.5$. This finding can be directly
related to the acoustic emission results if one assumes that a
fixed fraction of the work done by the external forces during an
elongation jump is released in the form of acoustic energy.

Theoretically, the occurrence of intermittent deformation bursts
has been modelled using both discrete dislocation simulations (see
e.g. Miguel et al. 2001) and different types of continuum models
(Koslowski et al. 2004, Zaiser and Moretti 2005). In the latter
case, the emergence of scale-free avalanche dynamics is related to
a depinning-like transition between an elastic and a plastically
deforming phase (`yielding transition'). A comprehensive overview
of the experimental and theoretical findings has been given by
Zaiser (2006).

While the temporal characteristics of plastic flow in single
crystals of various materials have received a great deal of
attention, much less has been done regarding the {\em spatial}
aspects of plastic deformation. Weiss and Marsan (2004) used
spatial triangulation to locate acoustic emission sources in
plastically deforming ice single crystals and found indications of
a fractal pattern of deformation loci with a fractal dimension of
about 2.5. Several investigators (Zaiser et al. 2004, Wouters et
al., 2005, 2006) studied the deformation-induced surface
morphology of polycrystalline specimens and observed self-affine
surface patterns with roughness exponents $\zeta = 0.7 \dots 0.9$
and an upper correlation length proportional to the grain size of
the polycrystalline aggregates.

However, a very straightforward question regarding a possible link
between temporal intermittency and spatial patterning of plastic
deformation has to our knowledge never been addressed. Plastic
deformation by crystallographic slip produces surface steps (in
dislocation language: passage of a dislocation leaves a monoatomic
step on the surface). It has been known for decades that these
steps are not evenly distributed but bundled together in so-called
slip lines or slip bands (for an overview see e.g. Neuh\"auser
1983). A conjecture that comes almost naturally to a
metallurgist's mind is that the emergence of each slip line or
slip band might be directly related to a single temporal
deformation burst - in other words, that the surface steps result
from deformation events that are localized both in space and in
time. In this case, the surface step pattern should be
intermittent in very much the same sense as the pattern of
deformation steps on the strain vs. time curves, and it should be
possible to match the respective statistics. The present letter is
exploring this conjecture.

Any investigation of the statistics of deformation-induced surface
steps hinges on the capability of preparing surfaces with a low
initial density of surface steps. Accordingly, in the present
investigation we use an alkali halide (KCl) single crystal
specimen where atomically almost flat surfaces can be prepared by
cleavage.

\section{Experimental}

\begin{figure}[tb]
\vspace*{.5cm}
\centerline{\epsfig{file=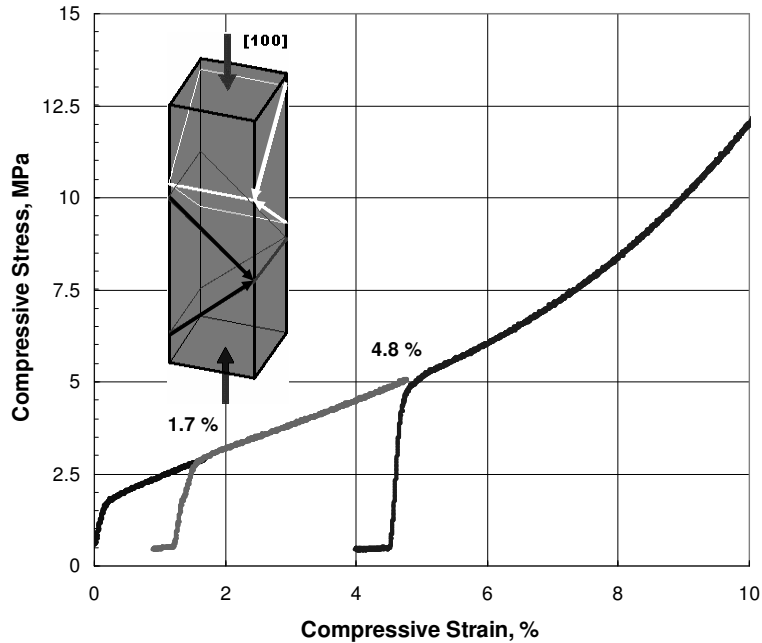,width=10cm,clip=!}}
\caption{Stress-strain curve showing the deformation protocol used
in the experiment. Surface maps were taken prior to deformation as
well as at 1.7 \%, 4.8 \% and 10\% compressive strain. Inset:
Deformation geometry showing the primary and orthogonal slip
system (slip dicrections under compression indicated by black
arrows) and the oblique slip systems (slip directions indicated by
white arrows).} \label{stressstrain}
\end{figure}

A KCl compression specimen with \{100\}-oriented faces and
dimensions of about $3.5\times3.5\times16$ mm$^3$ was prepared by
cleavage from a large commercial `optical grade' single crystal.
KCl cleaves readily along $\{100\}$ planes resulting in nearly
atomically flat surfaces. The specimen was deformed by compression
in successive steps in a standard testing machine, Instron 3360.
Tests were done in compression at room temperature, at a crosshead
rate of 0.002 mm/s corresponding to a nominal strain rate of $1
\times 10^{-4}$ s$^{-1}$. The specimen exhibits a typical
three-stage hardening curve with small constant hardening during
Stage I, an increased hardening rate in Stage II, and a decreasing
hardening rate in Stage III. The initial stages of the hardening
curve (Stage I and onset of Stage II) are seen in Figure
\ref{stressstrain}. In the specimen considered in the present
study, initial symmetry breaking let to the initiation of plastic
flow on one of four equivalent slip systems ($\{1\bar{1}0\}\langle
110 \rangle$, henceforth called the primary system) and
deformation then proceeded on this system throughout Stage I. At
the onset of Stage II, other slip systems (the so-called
orthogonal slip system and possibly the so-called oblique slip
systems) became active. For illustration of the slip geometry see
the inset in Figure \ref{stressstrain}.

Surface topography measurements were done with a scanning white
light interferometer (SWLI), Zygo NewView 100, with a lateral
resolution of about 0.5 $\mu$m and a vertical resolution of 0.3
nm. After examining the as-cleaved specimen, areas of $1 \times
0.36$ mm$^2$ were marked out on two adjacent side faces of the
specimen. These areas were chosen such as to exhibit as few as
possible large cleavage steps, and the surface topography was
recorded. Compression was then carried out in successive steps
with surface maps taken at $1.7\%, 4.8\%$ and $10\%$ compressive
strain, corresponding to the beginning and end of hardening Stage
I, and to hardening Stage II, respectively.

\begin{figure}[tb]
\vspace*{.5cm}
\centerline{\epsfig{file=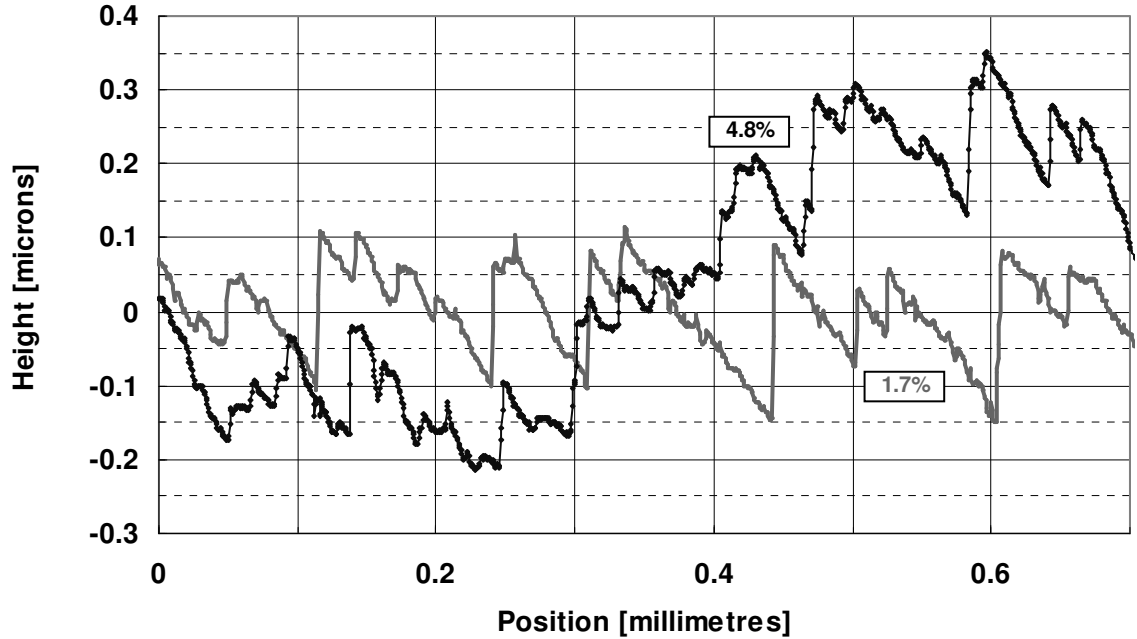,width=15cm,clip=!}}
\caption{Surface profiles of a KCl single crystal deformed in
compression along the [100] axis to compressive strains $\epsilon
= 1.7\%$ and $\epsilon = 4.8\%$.} \label{profiles}
\end{figure}

Profiles were taken after each deformation step on the specimen
face where slip steps from the primary and (in Stage II)
orthogonal slip systems emerged. The area coordinates on the
sample were fixed relatively to one of its corners allowing, in
combination with topographical characteristics such as large
cleavage steps, fairly precise location of the previously measured
area. For each deformation step, five 2D-profiles were extracted
parallel to the compression axis (i.e. orthogonal to the slip
lines). Profile lengths ranged between 0.7 and 1 mm, with a
sampling step of 0.55 $\mu$m. Surface profiles obtained after
deformation to compressive strains of $\epsilon = 1.7\%$ and
$\epsilon = 4.8\%$ are shown in Figure \ref{profiles}. At
$\epsilon = 10\%$ the specimen surface exhibits significant
macroscopic curvature as the specimen assumes a barrel shape. The
corresponding profile is not shown for the simple reason that it
does not fit into the same figure.

\section{Results and Discussion}

From the SWLI profiles, statistics of step heights and step
spacings (width of slip terraces) were determined. To this end,
the local slope of the profile was evaluated in terms of the
height difference of adjacent data points, a slip step was defined
as a compact interval of positive slope, and accordingly a slip
terrace as a compact interval of negative slope. Note that the
profiles are recorded such as to have zero average slope. If
deformation proceeds by slip on a single slip system, this implies
that any interval over which the local strain exceeds the average
strain has positive slope and, conversely, any interval over which
the local strain falls below the average strain has negative
slope. Hence, our definition of the slip steps and terraces
implicitly uses the average strain as a threshold value.

\begin{figure}[tb]
\vspace*{.5cm}
\centerline{\epsfig{file=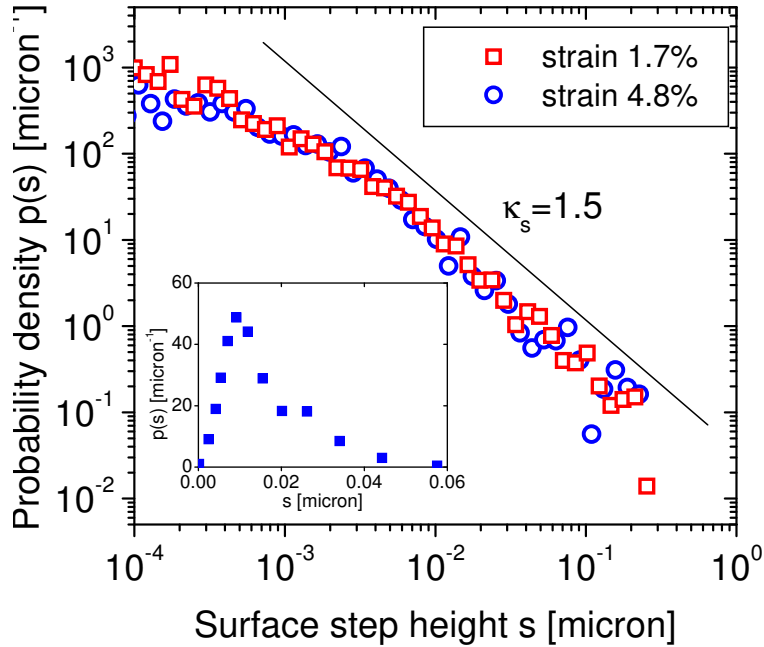,width=10cm,clip=!}}
\caption{Distribution of surface step heights after plastic
deformation; inset: cleavage steps on initial as-cleaved surface;
squares: total compressive strain 1.7 \%; circles: total
compressive strain 4.8 \%; each distribution has been determined
from 5 parallel profiles.} \label{stepdist}
\end{figure}

The initial as-cleaved surface exhibited cleavage steps with a
typical spacing of 30 $\mu$m. The probability density function of
the step heights is shown in the inset of Figure \ref{stepdist}.
It exhibits a peak at a typical step height of 10 nm, while the
mean step height is slightly larger.

After even a small degree of plastic deformation, the initial
cleavage steps are obliterated by the larger and more dense
deformation-induced slip steps. This can be seen from the
distributions of surface step heights that are shown in Figure
\ref{stepdist} for strains of $1.7\%$ and $4.8\%$. The surface
step heights now exhibit a scale-free distribution $p(s) \propto
s^{- \kappa_s}$ where $\kappa_s \approx 1.5$ for both $1.7\%$ and
$4.8\%$ strain. Scaling is observed above a slip step height of
about 2 nm (corresponding to about 6 dislocations leaving the
surface) and extends over two orders of magnitude in step height
without indications of an upper cut-off. The peak at a
characteristic step height of 10 nm, which was the signal feature
of the height distribution of cleavage steps, has disappeared
completely. Using a higher threshold value for the local slope in
the definition of a 'slip step' eliminates some of the smaller
slip steps but leaves the scaling regime virtually unchanged.

\begin{figure}[tb]
\vspace*{.5cm}
\centerline{\epsfig{file=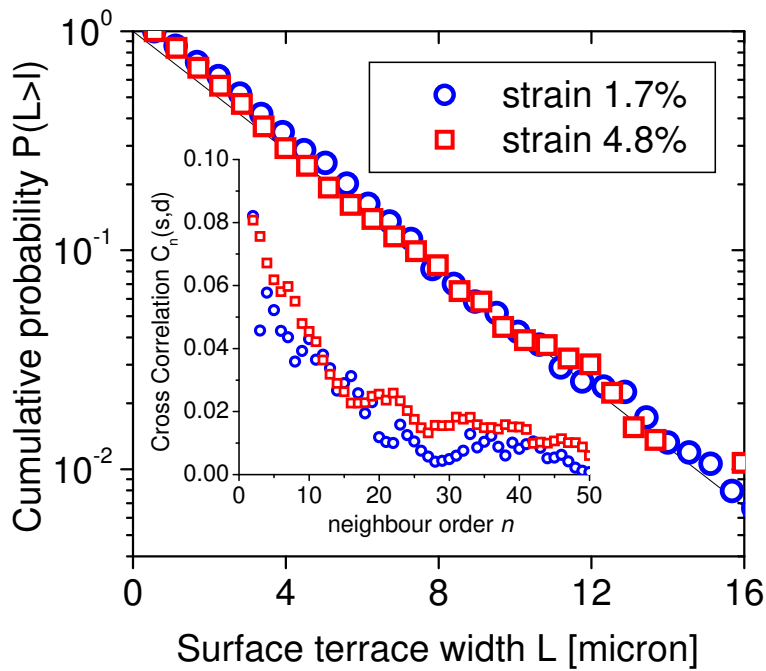,width=10cm,clip=!}}
\caption{Distribution of surface terrace widths (note the
single-logarithmic scale); squares: total compressive strain 1.7
\%; Circles: total compressive strain 4.8 \%; each distribution
has been determined from 5 parallel profiles. Inset: correlation
between height and $n$-th neighbor spacing of slip steps, as a
function of the neighbor order $n$} \label{terracedist}
\end{figure}

Cumulative probability distributions of slip terrace widths are
plotted in Figure \ref{terracedist}. These are quite different
from the slip step height distributions: The slip terrace widths
are exponentially distributed. Nevertheless, slip step heights and
slip step spacings turn out to be correlated. We have evaluated
the correlation functions $C_n(s,d_n)$ between the size $s$ of a
burst and the spacing $d_n$ of its $n^{\rm th}$ neighbors. To this
end, we enumerate the bursts along the profile and define
$d_n(x_i)$ (the $n$th neighbor spacing of the $i$th burst) as
$d_n(x_i) := x_{i+n} - x_{i-n}$. The correlation function
$C_n(s,d_n)$ is then defined as
$$
C_n(s,d_n) = \frac{\langle s_i d_n(x_i)\rangle}{langle s_i \rangle
\langle d_n(x_i) \rangle} - 1\PP
$$
This is plotted as a function
of $n$ in the insert of Figure \ref{terracedist}. One observes a
positive correlation between slip step height and slip step
spacing: Larger slip steps tend to be surrounded by larger slip
terraces and vice versa. This correlation can be detected up to
$n=30$, corresponding to a characteristic spacing $\langle d_n
\rangle$ of about 200 $\mu$m. The available statistics does,
however, not allow us to decide whether or not we are dealing with
a power-law decay.

The profiles obtained at 10\% strain (i.e., in hardening stage II)
are quite different from those obtained in hardening stage I. In
hardening stage II, multiple slip systems (the orthogonal and
possibly the oblique slip systems) become active. As a
consequence, the specimen surface develops significant macroscopic
curvature as the specimen assumes a barrel shape. This change in
shape has been corrected by subtracting a least-square fit 3rd or
4th order polynomial from the surface profiles before determining
the statistics of surface steps. Simultaneous activity of the
primary and orthogonal slip systems also implies that it becomes
quite impossible to distinguish between slip steps and slip
terraces, since slip steps created by the primary and orthogonal
systems have opposite `signs': On a crystal face where an edge
dislocation of the primary system creates an upward step of half a
lattice constant, an edge dislocation of the orthogonal system
creates a downward step of the same height. Accordingly, on the
profiles taken at 10 \% strain, upward and downward surface steps
obey similar statistics. The distributions are characterized by an
exponential decay at large step sizes, and no trace of power-law
statistics can be found.

The present findings regarding the slip step statistics in
hardening stage I (single slip) allow for a simple and
straightforward interpretation of the spatial `signature' of slip
events. The distribution of slip step heights is scale free with
an exponent of -1.5. This is the same exponent that has been found
for distribution of elongation increments observed by Dimiduk et
al. (2006) and the distribution of acoustic emission energies as
reported, e.g., by Miguel et al. (2001), indicating a one-to-one
correspondence between strain bursts and the emergence of discrete
slip steps.

It is interesting to compare the present findings with
investigations of the global surface morphology of plastically
deformed KCl single crystals as reported by Nadgorny et al. (2006)
and Zaiser (2006), and with the results of theoretical simulations
(Zaiser and Moretti 2005, Zaiser 2006). In these investigations,
non-trivial scaling of the height-height correlation function of
the surface profiles $h(x)$ was reported: $\langle |h(x) -
h(x+l)|\rangle \propto l^{\zeta}$ where $\zeta\approx 0.7$. Such
scaling with an exponent $\zeta > 0.5$ can be either due to the
presence of long range correlations in the surface strain
$\gamma(x) = \d h/\d x$, $\langle \gamma(x)\gamma(x+l)\rangle
\propto l^{2\zeta - 2}$ (self-affine scaling), or due to the
presence of fat tails in the distribution $p(s)$ of surface step
heights. Specifically, if surface steps are spatially uncorrelated
but characterized by a probability density $p(s)$ with a tail
$p(s) \propto s^{-\kappa_s}$, then for $\alpha = \kappa_s - 1 < 2$
the sum $h(x) = \sum_{x_i < x} s_i$ will scale like
$x^{1/\alpha}$, i.e., $\zeta = 1/\alpha$. On large scales, such a
profile would correspond to a Levy flight.

In the present case, we find a heavy-tailed distribution of
surface steps but the surface profiles cannot be envisaged as Levy
flights. Our value $\kappa_s = 1.5$ corresponds to $\alpha = 0.5$,
hence a sequence of uncorrelated slip steps drawn from this
distribution would lead to a large-scale surface profile which
might be envisaged as a Levy flight with $\alpha = 0.5, \zeta =
2$. In the actual profiles, however, the effect of the heavy tails
is partly offset by the long-range anti-correlation that arises
from the fact that larger slip steps are surrounded by larger slip
terraces and vice versa. The profiles cannot be interpreted as
self-affine, and they are not Levy flights either.

Further and more systematic investigations, and better statistics,
are clearly needed to understand the roughening of plastically
deformed metal surfaces. In the present investigation, the lateral
resolution along the surface achieved by SWLI is above the average
slip step spacing such that individual slip steps may get 'lost'
in the statistics. Further investigations should therefore include
techniques with higher lateral resolution such as atomic force
microscopy (AFM). Using AFM, it may also be possible to decide
upon the slip step statistics in higher deformation stages, where
optical data suffer greatly from lack of lateral resolution, both
in view of the smaller slip step spacings and in view of the
superposition of slip steps pertaining to multiple slip systems.

\section*{Acknowledgements}
Financial support by the European Commission under RTN/SizeDepEn
HPRN-CT 2002-00198 and of EPSRC under Grant No. EP/E029825 is
gratefully acknowledged.

\section*{References}

Dimiduk, D.M., Woodward, C., LeSar, R., Uchic, M.D., 2006.
Scale-Free Intermittent Flow in Crystal Plasticity, Science 26,
1188-1190.

Koslowski, M., LeSar, R., Thomson, R., 2004. Avalanches and
scaling in plastic deformation, Phys. Rev. Letters 93, 125502.

Miguel, M.C., Vespignani, A., Zapperi, S., Weiss, J., Grasso,
J.-R., 2001. Intermittent dislocation flow in viscoplastic
deformation. Nature 410, 667-671.

Nadgorny, E.M., Schwerdtfeger, J., Madani-Grasset, F., Koutsos,
V., Aifantis, E.C., and Zaiser, M., 2006. Evolution of self-affine
surface roughness in plastically deforming KCl single crystals.
In: A. El-Azab, S. Dattagupta, S.B. Krupanidhi, S. Noronha, S.S.
Shivashankar and M. Zaiser (Eds.), Proc. Int. Conf. on Statistical
Mechanics of Plasticity and Related Instabilities, Proceedings of
Science, PoS (SMPRI2005) 012.

Neuh\"auser, H., 1984. Slip-line formation and collective
dislocation motion. In: F.R.N. Nabarro (Ed.), Dislocations in
Solids, Vol. 4, North-Holland, Amsterdam, pp. 319-440.

Richeton, T., Dobron, P., Chmelik, F., Weiss, J., and Louchet, F.,
2006. On the critical character of plasticity in metallic single
crystals. Mater. Sci. Engng. A 424, 190-195.

Weiss, J., Grasso, J.-R., 1997. Acoustic emission in Single
Crystals of Ice. J. Phys. Chem. B 101, 6113-6117.

Weiss, J., Marsan, D., 2003. Three-dimensional mapping of
dislocation avalanches: Clustering and space/time coupling.
Science 299, 89-92.

Wouters, O., Vellinga, W.P., Van Tijum, R., De Hosson, J.T.M.,
2005. On the evolution of surface roughness during deformation of
polycrystalline aluminum alloys. Acta Mater. 53, 4043-4050.

Wouters, O., Vellinga, W.P., Van Tijum, R., De Hosson, J.T.M.,
2006. Effects of crystal structure and grain orientation on the
roughness of deformed polycrystalline metals. Acta Mater. 54,
2813-2821.

Zaiser, M., Madani, F., Koutsos, V., Aifantis E.C., 2004.
Self-affine surface morphology of plastically deformed metals.
Phys. Rev. Letters 93, 195507.

Zaiser, M., Moretti, P., 2005. Fluctuation phenomena in crystal
plasticity - a continuum model. J. Stat. Mech, P08004.

Zaiser, M., 2006. Scale invariance in plastic flow of crystalline
solids. Adv. Physics 55, 185-245.

\end{document}